# FANTAPPIÉ-ARCIDIACONO THEORY OF RELATIVITY VERSUS RECENT COSMOLOGICAL EVIDENCES : A PRELIMINARY COMPARISON

Leonardo Chiatti

ASL VT Medical Physics Laboratory
Via San Lorenzo 101
01100 Viterbo
Italy

fisica1.san@asl.vt.it

**Abstract**

Notwithstanding the Fantappié-Arcidiacono theory of projective relativity was introduced more than half a century ago, its observational confirmations in cosmology (the only research field where its predictions differ from those of the Einsteinian relativity) are still missing. In line of principle, this theory may be proposed as a valid alternative to the current views assuming the dominance of dark matter and inflationary scenarios.
In this work, the relativistic transformation of the Poynting vector associated with the reception of electromagnetic waves emitted by astronomical objects is derived in the context of the special version of the theory. On the basis of this result, and some heuristic assumptions, two recent collections of observational data are analyzed : the *m-z* relation for type Ia supernovae (SNLS, SCP collaborations) and the log *N* – log *S* relation obtained from the FIRST survey of radio sources at 1.4 GHz. From the first analysis, values are derived for the current density of matter in the universe and the cosmological constant that are of the same order of magnitude as those obtained from the most recent conventional evaluations. The second analysis results in an evolutionary trend of number of sources as a function of *z* that is in qualitative agreement with that obtained from more conventional analyses. Therefore it can be concluded, as a preliminary result, that the application of the theory to the study of cosmological processes leads to results which not substantially differ from these currently accepted. However, in order to obtain a more reliable comparison with observations, a solution is needed for the gravitational equations in the general version of the theory.

## 1. Introduction

The theory of projective relativity has had a rather peculiar fate among all the theories of physics : initially proposed in 1954 [1,2] by the famous Italian mathematician Fantappié (1901-1956) and subsequently worked in detail by Fantappié himself and his successors, in particular Arcidiacono (1927-1997) [3-15], it has never been subjected to experimental verification. This condition of incompleteness is due to technical as well as historical reasons, which we will attempt to summarize briefly.

From a technical point of view, the projective relativity (in the sense of Fantappié-Arcidiacono) effectively consists of two distinct theories : a projective special relativity (PSR) and a projective general relativity (PGR) [16-18]. The relationship between PSR and PGR is the same that exists between Einsteinian special relativity (SR) and general relativity (GR). PSR is a generalization of SR obtained by requiring the laws of physics to be invariant respect to the de Sitter group, instead of the Poincaré group. When the radius of the universe $r$ tends to infinity, the de Sitter group degenerates into the Poincaré group and PSR becomes the usual SR.

The spacetime geometry of PSR is clearly holonomic; its non-holonomic extension leads to PGR, where we have the following five-dimensional gravitational equations (of Arcidiacono) that generalize the four-dimensional Einstein equations :

$$R_{AB} - \frac{1}{2} R g_{AB} + \Lambda g_{AB} = \chi T_{AB} \qquad (1.1)$$

[where the cosmological term has been added]. In the limit $r \to \infty$, these equations become the conventional Einsteinian equations of GR. Therefore, the PSR predictions differ from those of SR only when the size of the considered systems is not negligible with respect to $r$, or the time scale of the considered processes is not negligible with respect to $t_0 = r/c$, being $c$ the speed of light in the vacuum. The same conclusion holds for the divergence between the predictions of PGR and those of GR. In summary, we can thus say that the projective relativity predictions differ from those of ordinary Einsteinian relativity only at the cosmological scale. The experimental verifications of the projective relativity should therefore consist, in reality, in observational verifications in the cosmological context.

However, from an historical point of view, the theory was created and developed in mathematical or mathematical physics-related environments, wherein the interest for its application to cosmology or extragalactic astrophysics was not a priority. Thus the most interesting subjects for a comparison with the data coming from observational cosmology have not yet been dealt with. For example, there still lacks a classification of the solutions of equations (1.1) under the assumption of the cosmological principle, i.e., the analogue of the Fridman classification of the cosmological models for GR. Thus it is not currently possible to use the equations (1.1) in order to analyze the type Ia supernovae data and obtain deductions on the cosmological model or the extragalactic distance scale, as instead occurs in current relativistic cosmology. PSR is instead completely developed, but it provides a relation between redshift and distance that does not take into account gravitation or the cosmological term [16-18] and that therefore is not directly applicable to observational data. In a later section, we will limit ourselves to the Newtonian approximation of PGR. We will see that the analysis of some sets of recent observational data in this framework seems to lead to conclusions that are not very dissimilar to those of current relativistic cosmology. One could therefore ask why we should choose projective relativity rather that the more usual Einsteinian relativity. As Arcidiacono himself realized, projective relativity naturally resolves many problems that afflict the standard big bang. First of all, PSR seems to incorporate a sort of perfect cosmological principle : all observers, in any epoch, are located at the same chronological distance $t_0$ from the big bang (in the past) as well as the big crunch (in the future). Each observer, moving along his time line, remains at the same distance from the big bang and from the big crunch exactly like a ship, sailing across the ocean, reaches faraway locations while remaining at the same distance from the horizon. The big bang and the big crunch are therefore limiting surfaces in the same sense in which the light cone is, and not singularities that an observer can experience "here and now"; this resolves the problem of the singularities. Different inertial observers are connected by coordinate transformations whose set is isomorphic to the rotations of the five-dimensional sphere around its centre. The radius of this sphere is $r$ and its value has no relationship with the distribution of matter or energy over the spacetime. In this sense the global curvature of the space (to be understood here in terms of the intrinsic geometry of the spacetime, i.e., as a parameter that connects the observations made by different observers, rather than in the extrinsic sense of an effective curvature of the spacetime in a "fifth dimension") is a fundamental constant of nature totally independent from the presence and distribution of matter-energy. Each observer coordinates the events in a four-dimensional spacetime (Castelnuovo spacetime [19-21] in the case of PSR) which is, in any case, flat. This resolves the flatness problem, removing the constraint $\Omega = 1$ and liberating the community of research workers from the nerve-racking obligation to search for "dark mass" or "missing mass"[1]. This does not prevent, naturally, the existence of a share of ordinary dark matter.

Given an observer O' located on the past light cone of a second observer O, the edge of his light cone inside the light cone of O (and therefore not connected with O through light signals) depends on the relative position of O and O'. When O' approaches the big bang, the opening of his light cone (in the reference frame with O as the origin) grows indefinitely (Fig. 1), which resolves the problem of the initial homogeneity and isotropy without any need to postulate inflationary mechanisms [17].

Finally, as recently commented by Licata [22-23], PSR seems to be an excellent starting point for rethinking the entire problem of the foundation of quantum cosmology, also providing a natural interpretation for the Hartle-Hawking solution.

The structure of this work is articulated as follows. The second section presents the fundamentals of the covariant formulation of electrodynamics in the context of PSR. In the third section, we derive the transformation of the Poynting vector associated with the reception of electromagnetic waves emitted by astronomical objects. In the fourth section the PSR distance-red shift relation is used to analyze the log $N$ – log $S$ radio sources counts at 1.4 GHz (non-selective FIRST survey, first set of data on 1550 square degrees of the northern sky). The problem of matter evolution is briefly discussed in the fifth section. In the last section a Newtonian approximation of PGR cosmology is considered and recent data on type Ia supernovae (SNLS, SCP collaborations) are analysed according to it.

## 2. Cosmic electromagnetism

In their conventional form, the Maxwell equations of electromagnetism and the expression of the Lorentz force acting on charges and currents are covariant with respect to the Poincaré group. In PSR these relations must be appropriately generalized so as to be covariant with respect to the Fantappié group (in practice, the projective version of the de Sitter group). Such a generalization gives the following expressions, where the notations have the usual meaning [16-18] :

$$\mathrm{div}\,\mathbf{E} = \rho \qquad \mathrm{div}\,\mathbf{H} + \bar{\partial}_0 C_0 = 0$$

$$\mathrm{curl}\,\mathbf{E} + \bar{\partial}_4 \mathbf{H} - \bar{\partial}_0 \mathbf{C} = 0 \qquad (2.1)$$

$$\mathrm{curl}\,\mathbf{H} - \bar{\partial}_4 \mathbf{E} = \mathbf{j}$$

$$\mathrm{div}\,\mathbf{C} + \bar{\partial}_4 C_0 = s \qquad \mathrm{curl}\,\mathbf{C} - \bar{\partial}_0 \mathbf{E} = 2\boldsymbol{\omega}$$

$$\mathrm{grad}\,C_0 + \bar{\partial}_4 \mathbf{C} - \bar{\partial}_0 \mathbf{H} = \mathbf{a} \qquad (2.2)$$

---

[1] According to the current use, we indicate as $\Omega$ the ratio between the actual energy-matter density and its critical value.

$$\mathbf{F} = (\rho \mathbf{E} + \mathbf{j} \times \mathbf{H}) - (\mathbf{a} C_0 + 2\boldsymbol{\omega} \times \mathbf{C})$$

$$F_4 = i(\mathbf{j} \cdot \mathbf{E} - \mathbf{a} \cdot \mathbf{C}) \qquad F_0 = \mathbf{a} \cdot \mathbf{H} - 2\boldsymbol{\omega} \cdot \mathbf{E} \qquad (2.3)$$

These results require a comment. As can be seen, in addition to the usual electric **E** and magnetic **H** fields appear two new fields : one a scalar ($C_0$), and the other a vector (**C**). In the corresponding generalization of the ponderomotive force (2.3), these fields are coupled with the currents **a** ed **ω**. For a correct reading of the equations, there should be kept in mind that the derivatives that appear are projective derivatives, and not conventional partial derivatives. We have used the notation of Arcidiacono, so the index 4 refers to the time axis, the index 0 to the fifth coordinate and the indices 1, 2, 3 to the space coordinates.

The equations (2.1) take the name of "Maxwellian" equations, while the equations (2.2) are known as "non-Maxwellian" equations. It is not difficult to understand the genesis of this second group of equations[2]. In an infinite universe, the photon has a null rest mass $M$ and it moves at the speed of light; its Compton wavelength is thus $\lambda_C = \hbar/Mc = \infty$. But if the radius of the universe $r$ is finite, we cannot have $\lambda_C \gg r$ and thus $\lambda_C = \lambda_{C\,max} \approx r$. It follows that in a certain sense the photon acquires a finite mass $\approx \hbar/rc$. It therefore acquires a longitudinal component and this is the reason for the appearance of the fields (**C**, $C_0$). The equations (2.2) formally coincide, for $r = \infty$, with the non-Maxwellian photon equations obtained by de Broglie with the fusion method on the Minkowski spacetime, when the rest photon mass is negligible; these latter equations, in fact, describe precisely the longitudinal component of the photon [24,25].

For $r \to \infty$, the equations (2.1), (2.2) are not more coupled and we must recover the ordinary Maxwell equations that were the starting point of the generalization; therefore the disappearance of non-Maxwellian equations is requested. In turn, the necessary disappearance of the non-Maxwellian fields implies the inexistence of distinct sources of these fields; in other words, there must be identically $s = \mathbf{a} = \boldsymbol{\omega} = 0$. The only origin of the fields (**C**, $C_0$) is therefore the relativistic transformation (in the framework of PSR) of the ordinary fields (**E**, **H**) associated with remote charges and currents. Let's consider, as an example, the equations :

$$C_0' = C_0 \; ; \qquad \mathbf{C}' = (\mathbf{C} - \gamma\,\mathbf{H}) / (1 - \gamma^2)^{1/2} \; ;$$

$$\qquad\qquad\qquad\qquad\qquad\qquad\qquad\qquad\qquad\qquad (2.4)$$

$$\mathbf{E}' = \mathbf{E} \; ; \qquad \mathbf{H}' = (\mathbf{H} + \gamma\,\mathbf{C}) / (1 - \gamma^2)^{1/2} \; .$$

which express the transformation of the fields due to a time translation of length $T_0$. We have set $\gamma = T_0/t_0$, where $t_0 = r/c$ and $r$ is the radius of curvature of the de Sitter spacetime. It is immediately evident that at the Minkowskian limit these relations become invariances. Given the absence of local sources of non-Maxwellian fields, these latter must be null here and now. Therefore $C_0 = 0$, **C** = 0 and :

---

[2] We will not follow, in this work, the mistaken conviction of Arcidiacono. He believed that the fields **C**, $C_0$ were associated with a fluid medium and that the equations (2.1), (2.2) therefore described a unification of electromagnetism with hydrodynamics! Generalizing the Maxwell equations to more than 5 dimensions, he claimed to have obtained a unified theory of electromagnetic, gravitational, hydrodynamic, etc. fields while he was actually dealing simply with the components of the generalized electromagnetic field.

$$C_0' = 0 \ ; \qquad C' = -\gamma \mathbf{H} / (1 - \gamma^2)^{1/2} \ ;$$

$$\mathbf{E}' = \mathbf{E} \ ; \qquad \mathbf{H}' = \mathbf{H} / (1 - \gamma^2)^{1/2} \ . \qquad (2.5)$$

The second relation expresses the appearance of a field **C'** as transformation of the magnetic field **H** at cosmological distances. Thus, if we observe the effects of a very distant magnetic field in the past (let's say five billion years ago) these effects will be modified due to an increase in the field strength by a factor of $1/(1-\gamma^2)^{1/2}$, and due to the simultaneous appearance of a field **C'** associated with the transformation of **H**.

How we can see by substituting the identities $s = 0$, $\mathbf{a} = 0$, $\boldsymbol{\omega} = 0$[3] into the equations (2.3), the fields (**C**, $C_0$) are not coupled with the matter and thus, in particular, they do not contribute to the luminosity of a remote astronomical object. More generally, the energy associated with the non-Maxwellian fields will appear, in the energy balance of the electromagnetic field, as "missing" energy which is not instrumentally detectable in a direct manner.

Obviously, if the relativistic transformations of the fields **E** ed **H** that alter the luminosity of extragalactic objects are not taken into account, the evaluations of the extragalactic scale distance using standard candles can be affected by significant systematic errors. A study of the problem requires the evaluation of how the Poynting vector transforms passing from a reference frame with origin at the remote source to a new reference frame with origin at the observation pointevent.

## 3. Relativistic transformation of the Poynting vector (in the framework of PSR)

PSR is a five-dimensional theory, hence we have two Poynting vectors [16-18] :

$$T_{\alpha 4} = i(C_0 \mathbf{C} + \mathbf{E} \times \mathbf{H}) \qquad T_{\alpha 0} = C_0 \mathbf{H} + \mathbf{E} \times \mathbf{C} \qquad (3.1)$$

The physical interpretation of these vectors is not difficult : their fluxes through a closed hypersurface containing the entire three-dimensional space provides the components of the momentum of the electromagnetic field along the time axis and along the fifth axis respectively. The component of the momentum along the fifth axis does not have direct physical effects, so we should concentrate our attention on the first of the equations (3.1). This equation generalizes the ordinary Poynting vector including the flux of energy associated with the longitudinal component of the electromagnetic waves. In the Einsteinian limit $r \to \infty$ the first of the equations (3.1) reduces to the ordinary Poynting vector of normal electromagnetic theory, while the second equation transforms into the identity $0 = 0$.

We then take as the primed reference frame that having the origin at the pointevent of emission by the astronomical object which emits electromagnetic waves, and as the unprimed reference frame that having the origin at the observation pointevent. The transformation of coordinates connecting the two reference frames consists in the product of a time translation of the origin with parameter $T_0$, a space translation of the origin with parameter $T$ and a boost with speed $V$. We will set, following the customary notation in the literature, $\alpha = T/r$, $\beta = V/c$, $\gamma = T_0/t_0$. The general transformation rules of the electromagnetic field components are then [26] :

---

[3] These equations are the duals of the relations expressing the inexistence of the magnetic charge and magnetic current. In fact, the magnetic field itself derives from a relativistic transformation of the electric field in the context of conventional SR : it is generated by electric fields *in motion*, like the fields C derive instead from the *distant* magnetic fields. All this is due to the simultaneous existence in PSR of two fundamental constants: the *speed c* and the *distance r*.

$$E_1' = E_1$$

$$E_2' = a_{11} E_2 + a_{14} H_3 + a_{15} C_3$$

$$E_3' = a_{11} E_3 + a_{14} H_2 + a_{15} C_2 \qquad (3.2)$$

$$H_1' = a_{51} C_0 + a_{54} C_1 + a_{55} H_1$$

$$H_2' = a_{41} E_3 + a_{44} H_2 + a_{45} C_2$$

$$H_3' = a_{41} E_2 + a_{44} H_3 + a_{45} C_3$$

$$C_0' = a_{11} C_0 + a_{14} C_1 + a_{15} H_1$$

$$C_1' = a_{41} C_0 + a_{44} C_1 + a_{45} H_1$$

$$C_2' = a_{51} E_3 + a_{54} H_2 + a_{55} C_2$$

$$C_3' = a_{51} E_2 + a_{54} H_3 + a_{55} C_3$$

Where:

$$B\, a_{11} = 1$$

$$B\, a_{41} = \beta$$

$$B\, a_{51} = \beta\gamma - \alpha$$

$$AB\, a_{14} = \beta + (\alpha - \beta\gamma)\gamma \qquad (3.3)$$

$$AB\, a_{44} = 1 + (\alpha - \beta\gamma)\alpha$$

$$AB\, a_{54} = \gamma - \alpha\beta$$

$$A\, a_{15} = \alpha$$

$$A\, a_{45} = \gamma$$

$$A\, a_{55} = 1$$

$$A^2 = 1 + \alpha^2 - \gamma^2$$

$$B^2 = 1 - \beta^2 + (\alpha - \beta\gamma)^2$$

We now have to specialize these general rules by inserting the special conditions of our problem. First of all, let us consider the Poynting vector incident on the observer as it appears in the unprimed reference frame. The requirement is that the fields incident on the observer consist of transversal spherical waves locally coincident with plane waves[4]. Under these conditions, the non-Maxwellian fields evaluated in the unprimed reference frame vanish :

$$C_0 = C_1 = C_2 = C_3 = 0 . \tag{3.4}$$

Furthermore, the fields **E** and **H** are perpendicular to one another and perpendicular to the direction of propagation; if we choose as the *x* axis (axis 1) that which joins the observer to the source, this axis will also be the axis of propagation. We can therefore take as the *y* axis (axis 2) the axis along which the electric field oscillates and as the *z* axis (axis 3) that along which the magnetic field oscillates. With these choices only the field components $E_2$ and $H_3$ survive, being :

$$E_1 = E_3 = H_1 = H_2 = 0 . \tag{3.5}$$

Moreover, there exist some relations between the transformation parameters. First of all, we note that in PSR the relation between the recession speed (that can also be superluminal) and redshift is expressed by $V = cz$, so we immediately obtain $\beta = z$. Furthermore, the source must necessarily be localized on the past light cone of the observer, so $\gamma = -\alpha$. The relation between $\alpha$ and $z$ provided by PSR is $\alpha = z/(1+z)$, so after all we have the substitutions:

$$\beta = z , \quad \gamma = -\alpha , \quad \alpha = z/(1+z) . \tag{3.6}$$

The application of the transformation rules (3.2), (3.3) then gives the Poynting vector in the reference frame of the source, evaluated at the observation pointevent. Recall that we are dealing with the plane waves approximation and that in PSR the spatial section is Euclidean; therefore the geometric attenuation factor (the inverse of the squared distance) must be taken into account even it does not appear in our calculation explicitly.
Proceeding with the calculation, we note that the only not vanishing primed components are $E_2'$, $H_3'$, $C_3'$, therefore the only not vanishing component of the Poynting vector is that along the axis 1; in the usual CGS units it becomes $S_1' = (c/4\pi)E_2'H_3'$. In the unprimed frame we have, in CGS units again, $S_1 = (c/4\pi)E^2$, where $E_2 = H_3 = E$. A simple calculation then gives :

$$\frac{S_1'}{S_1} = \frac{z}{B^2} + \frac{[z - \gamma^2(1+z)][1 + \gamma^2(1+z)]}{A^2 B^2} + \frac{[z - \gamma^2(1+z)]z + [1 + \gamma^2(1+z)]}{AB^2}$$

From the second of equations (3.6) we have $A^2 = 1$ and thus, with the usual choice of the sign of the fifth homogenous coordinate, $A = 1$. Substituting all the relations (3.6) into the expression for $B^2$ we then have $B^2 = 1$. Algebraically developing the relation obtained we derive the following expression :

$$\frac{S_1'}{S_1} = (1+z)^2 (1 - \alpha^4) \tag{3.7}$$

which constitutes the main result of this section. If the second factor on the right hand is temporarily neglected, we obtain the well known relation $S_1 = S_1'/(1+z)^2$ giving the attenuation of the absolute luminosity (defined in the reference frame of the source) due to the redshift. In the usual calculation of the source distance starting from its absolute magnitude (assumed to be known) and from its apparent magnitude measured in a given photometric system, this is the only kinematic factor taken into consideration. The equation (3.7) tells us however that the flux $S_l'/(1+z)^2$ is equal to $S_l(1-\alpha^4)$, not $S_l$. Since $(1-\alpha^4)$ is between 0 and 1, at equal absolute luminosity and redshift the apparent luminosity is greater than (and then the luminosity distance is less than) normally believed. In the case of standard candles, calibrated with sources relatively near Earth, the absolute magnitude is reasonably well known, hence it is the distance that is overestimated. The correct luminosity distance $x_{corr}$ is related to the luminosity distance $x_{uncorr}$, measured without taking into consideration the correction, through the relation :

$$x_{corr} = x_{uncorr} \sqrt{1-\alpha^4} = x_{uncorr} \sqrt{1-\left(\frac{x_{corr}}{r}\right)^4} \tag{3.8}$$

that can be written more usefully as the solution of the corresponding biquadratic equation :

$$x_{corr} = \frac{\sqrt{\sqrt{1+4\left(\frac{x_{uncorr}}{r}\right)^4}-1}}{\sqrt{2}\,\frac{x_{uncorr}}{r^2}} \tag{3.9}$$

---

[4] The spherical waves emitted from the source placed at the origin of the primed reference frame will be describable as superpositions of plane waves in that reference frame. The attenuation factor (3.7), derived for the plane waves in the primed reference frame, therefore also holds for the spherical waves derived from their superposition.

The equation (3.9) was inferred using the third equation of (3.6) that represents the relationship between redshift and distance in PSR. In a successive section, we will apply equation (3.9) to real observational data, assuming that it also holds outside the domain of application of PSR, i.e., in the more extensive domain of PGR.
The effect of the correction of the distances is clear : with the increasing redshift the correct distance asymptotically approaches the limit distance $r$, while the uncorrected distance increases beyond this limit[5].

## 4. Radio source counts

The projective relativity includes a sort of perfect cosmological principle, so the relation between evolutionary aspects and non-evolutionary aspects in the history of the universe needs of a clarification. Historically speaking, the radio source counts have been decisive in indicating the existence of evolutionary aspects [27]. We propose an examination of the integral counts along the following lines.
Under the hypothesis of an Euclidean spatial section (which is exact for PSR, while it remains to be verified in the ambit of PGR), the flux $S$ incident on the observer from an extragalactic radio source with luminosity $L_0$ and spectral index $\eta$ observed at the frequency $v$ is :

$$S = \frac{L_0 v^{-\eta}}{4\pi x^2 (1+z)^{1+\eta}} \tag{4.1}$$

---

[5] This is obvious in PSR, where the third relation (3.6) holds: for z → ∞ it becomes exactly $x_{corr} \to r$. If, as we have done, a broader range of applicability is assumed for eq. (3.9), the same result can be obtained by taking its limit for $x_{uncorr} \to \infty$.

where $x$ is the uncorrected luminosity distance of the source and $z$ is the redshift. If all the sources were identical, the flux of one of these would be greater than $S$ only if its luminosity distance were inferior to :

$$x = \sqrt{\frac{L_0 \nu^{-\eta}}{4\pi S (1+z)^{1+\eta}}} \quad .$$

Now the corrected luminosity distance $y$ is related to the uncorrected luminosity distance $x$ by eq. (3.8). The flux of the radio source is therefore greater than $S$ if it is inside the sphere with centre at the observer having the radius :

$$y = \sqrt{\frac{L_0 \nu^{-\eta} (1-\alpha^4)}{4\pi S (1+z)^{1+\eta}}} \quad ,$$

where $\alpha = y/r$. If the numerical density of radio sources $\delta$ is homogeneous, the number of radio sources contained in this sphere is :

$$N(S) = \frac{4}{3}\pi y^3 \delta = N_0(S) (1+z)^{-\frac{3}{2}(1+\eta)} (1-\alpha^4)^{\frac{3}{2}}$$

where the factor $N_0(S)$ is proportional to $S^{-3/2}$. Therefore an observational determination of the quantity $S^{3/2}N(S)$, where $N(S)$ is the number of sources with flux greater than $S$, must lead to the result :

$$S^{\frac{3}{2}} N(S) = K (1+z)^{-\frac{3}{2}(1+\eta)} (1-\alpha^4)^{\frac{3}{2}} \quad , \tag{4.2}$$

where $K$ is an instrumental constant. The second member (apart from the factor $K$) can be theoretically calculated up to a constant, while the first member can be determined by observations. The theoretical calculation of the second member starts from the value of $\alpha$. Given $\alpha$, we have $y = r\alpha$ and the PSR relation $\alpha = z/(1+z)$ gives $z$. Since both $\alpha$ and $z$ are known, the second member can now be calculated. Since both $y$ and $z$ are known, the flux $S$ can now be calculated by applying equation (4.1), where $x$ is replaced by $y$. This procedure is iterated over all the values of $\alpha$ from 0 to 1, thus deriving the function that relates the right hand of (4.2) to the flux $S$. Since the numerator of (4.1) is actually unknown, the flux is defined up to an arbitrary factor $K'$. If all the theoretical assumptions are correct, reporting on a double logarithmic plot the curves representing the two members of eq. (4.2) as a function of the flux, these curves must be parallel. The theoretical curve (right hand) is shifted with respect to the experimental curve (left hand) not only vertically (due to the presence of $K$), but also horizontally (due to the presence of $K'$). More generally, the deviation from parallelism will provide indications on the limits of validity of the assumed hypotheses.

We have applied this analysis strategy to the preliminary results of the FIRST survey [28]. These results refer to a scanning of 1550 dcg$^2$ of the northern sky between the galactic latitudes +28° and +42°, at the frequency of 1.4 GHz, with a census of all the sources present. It does not involve, therefore, any source selection by type, luminosity class or spectral index. In particular, table 2 of ref. [28] has been used, which provides the differential counts normalized between 1 mJy and 1 Jy. The comparison between observational data and theoretical data is shown in Fig. 2; the error bars on the experimental data are not visible because they are very small. The spectral index has been assumed to be equal to 0.75, but considerable modifications of this value (in the range 0.75–1) do not lead to substantial changes.

How one can see, the well known excess of radio sources at small fluxes is confirmed. This excess is believed to be the proof of an evolutionary trend of radio source number and luminosity, compatible with the hypothesis of an initial singularity. Thus it can be stated that, even when taking into account the PSR correction of the distances, there is an excess of radio sources at small fluxes that supports an evolutionary universe. As a result, the perfect cosmological principle implied by PSR must be questioned, as discussed in the next section.

**5. Steady state or evolutionary universe?**

It wouldn't be possible to proceed in this work without commenting on the problem of the evolution of matter; in fact, this problem assumes a particularly delicate connotation in projective relativity. The current abundances of the chemical elements derive from the evolutionary processes which occurred during the interval $t_0$ passed since the big bang. But, under the sort of perfect cosmological principle derived from PSR, each observer, in any epoch and position, is distant $t_0$ from the big bang. Therefore, the abundances of the chemical elements present here and now would have to be the same everywhere and always : the universe must not evolve chemically.

Naturally, this doesn't mean that the single objects of the universe (clusters, galaxies and, inside them, the stars) do not evolve chemically. The thermonuclear reactions produce heavier elements, in the end releasing them into the interstellar/intergalactic medium. So arises the question : how is it possible that the abundances of the elements remain, on average, unchanged despite this release? How is it possible that the single objects age but the matter considered on the large scale, on average, does not age?

Each observer, observing galaxies at ever-increasing $z$, finds that these cluster ever-more densely : the surface $x = r$ is an impassable barrier; thus in his reference frame the density of galaxies grows indefinitely with the increase in $z$ and with the approaching of $x$ to $r$. If the origin of this reference frame is time translated in the future, a part of the galaxies that were first diffused in the past light cone will be more densely clustered around the big bang singularity. However, new galaxies, first outside the past light cone, will enter inside it, so affecting the chemical environment of the observer. It is therefore plausible that, on average, the cosmic abundances of the elements measured by the observer do not change.

The hypothesis that the universe does not evolve chemically is in strong disagreement with the data concerning the blackbody background radiation, the abundances of the light elements, the radio source counts and many other evidences. Also considering anomalies such as the recent discovery of the quasar APM 08279+5255 at $z = 3.91$, with an abnormally high content of iron [29,30], the hypothesis of a stationary universe seems difficult to support nowadays.

One way to get around the problem, always remaining within the same theoretical framework, is to explicitly break a symmetry, postulating that the total number of galaxies (or better, of "material points" understood as elementary particles) present within the Cayley-Klein absolute $A^2 = 1 + \alpha^2 - \gamma^2 = 0$, $\alpha = x/r$, $\gamma = t/t_0$ is finite, although very large, and that their world lines do not touch the singularities. In this case moving the origin of the reference frame along its time line towards the big bang, we arrive, in the end, to not having any more material points in the past light cone, emerging from the singularity. We will call this particular position of the origin on the time line the "alpha point" of that line. Similarly, moving the origin of the reference frame along its time line towards the big crunch, we arrive, in the end, to not having any more material points in the future light cone, entering into the singularity. We will call this position of the origin on the time line the "omega point" of that line. Keep in mind that both the alpha point as well as the omega point, just like any other intermediate point, are located at the same distance, equal to $t_0$, from the two singularities.

Each time line joining the two singularities will have its own alpha point and its own omega point. A material point that travels that line will begin to undergo the action of the matter only after the alpha point and will stop acting on the matter after the omega point. Before the alpha point, it may act on the matter but not vice versa (there is future, but not past); after the omega point, it may undergo the action of the matter but not vice versa (there is past, but not future). In the interval between alpha and omega, it may both act on other points as well as undergo their influence (there is both past and future).

Naturally, the requirement that the ends of the world lines of all the "material points" do not intersect the singularities is equivalent to requiring that there is an authentic creative phase of emergence of the matter from the vacuum, and an authentic destructive phase of annihilation of the matter into the vacuum. The explanation of these phases (vacuum transitions ?), which would in any case occur in a tranquil flat spacetime with singularities distant $t_0$, except for PGR effects to be analyzed, would clearly be found outside projective relativity : it concerns a more general cosmological theory, perhaps a quantum cosmological theory. Concerning the intermediate history of the matter, it could then be more or less that which nuclear astrophysics has made us all accustomed to by now : baryon era, lepton era, decoupling of the radiation, evolution of the structures.

Having broken the symmetry of time homogeneity (which remains valid on the level of the laws, but not on the level of the conditions) there no longer exists a perfect cosmological principle, but there can still exist a cosmological principle in the current sense of the term[6]. Thus we would return to a framework no longer in conflict with that agreed upon today. The next section assumes the validity of the cosmological principle in the ordinary sense of the term.

## 6. Considerations on the extragalactic scale distance from the SNLS, SCP data

In PSR the recession speed of an extragalactic object located at the spatial distance $x$ from the observer, evaluated at the time $t$ (where $t = 0$ is the present of the observer) is expressed by $V = Hx/(1 + t/t_0)$, where $H = c/r = 1/t_0$ e $-t_0 \leq t \leq 0$. The "Hubble's constant" at the time $t$ is therefore expressed by $H(t) = H/(1+ t/t_0)$ and, if it originated from the variation of the scale factor $R(t)$, we would have :

$$\frac{\dot{R}(t)}{R(t)} = H(t) = \frac{H}{1+\dfrac{t}{t_0}} \qquad (6.1)$$

from which it immediately follows that $R(t) = k|1+t/t_0|$, where $k$ is an appropriate integration constant. If we set $R = 1$ at the current time, in other words at $t = 0$, we then obtain $R(t) = |1+t/t_0|$. We now translate the origin of the time axis, so as to count the time starting from the big bang instead of the present. This means performing the substitution $t \to t - t_0$, which transforms the function $R(t)$ into the new function $R_0(t) = t/t_0$, with $t_0 \geq t \geq 0$. The present time is now $t = t_0$, and $H(t)=H_0(t)=1/t$. The physical meaning of $R_0(t)$ is transparent : if the expansion of the universe unaffected by gravitation and the cosmological constant, which in PSR appears as an expression of the Fantappié group kinematic, derives instead from the time variation of a scale factor, this scale factor would be precisely $R_0(t)$.

The problem arises when one tries to generalize eq. (6.1) by including a genuine expansion of the space, represented by a scale function, together with the purely apparent, projective expansion; this problem leads to the PGR. The rigorous derivation of Hubble's law in this more general case presumes the solution of the gravitational equations (1.1) under the assumption of the cosmological principle. Since these solutions are not available in the current literature, we will handle the problem in Newtonian approximation, slightly modifying the original subject of Milne and McCrea [33]. Calculating the time $t$ starting from the big bang, the usual expression of Hubble's "constant" on the plane of contemporaneousness of the observer is modified in the following manner :

$$V(t) = K(t)x \qquad (6.2)$$

$$H(t) = K(t) - H_0(t) = \frac{\dot{R}(t)}{R(t)} \qquad (6.3)$$

---

[6] In other words, the perfect cosmological principle still holds for a pointevent respect to the null geodesics entering into it, while the usual cosmological principle holds for a real, physical observer which sees the Universe through the light rays emitted by other material bodies.

Where $x$ is the spatial distance from the observation point. For the cosmological principle, the density of matter $\rho(t)$ and the pressure $p(t)$ depend only on the cosmic time $t$. Applying the continuity equation for the cosmic fluid, we then have :

$$\frac{\partial \rho}{\partial t} + div(\rho \vec{v}) = \frac{\partial \rho}{\partial t} + div[\rho K(t)\vec{x}] = \frac{\partial \rho}{\partial t} + 3\rho K(t) = 0. \qquad (6.4)$$

Substituting eq. (6.3) in eq. (6.4), integrating and setting $R(t_0)=1$, we obtain :

$$\rho(t) = \frac{\rho(t_0)}{[R(t)(\frac{t}{t_0})]^3} \qquad (6.5)$$

This equation describes the variation of the density of matter over cosmic time. The force $F$ acting on a unit of mass of the cosmic fluid is expressed by the Euler equation :

$$\frac{D\vec{v}}{Dt} + \frac{grad(p)}{\rho} - \vec{F} = 0$$

The second term is null since $p$ depends only on $t$. Thus :

$$\frac{d}{dt}[K(t)\vec{x}] - \vec{F} = [K\vec{v} + \dot{K}\vec{x}] - \vec{F} = (K^2 + \dot{K})\vec{x} - \vec{F} = 0 .$$

If there are only gravitational forces and a cosmological term $\lambda$, the divergence of this expression becomes :

$$3(K^2 + \dot{K}) = -4\pi G\rho + \lambda$$

Where $G$ is the Newtonian gravitational constant. Substituting eq. (6.3) into this expression and rearranging the terms, one obtains :

$$\frac{\ddot{R}}{R} + 2H_0 \frac{\dot{R}}{R} = -\frac{4\pi G\rho}{3} + \frac{\lambda}{3}.$$

Multiplying the two members for $R^3$ and inserting eq. (6.5), we have :

$$R^2[\ddot{R} + 2H_0\dot{R}] + \frac{4\pi G\rho(t_0)}{3(\frac{t}{t_0})^3} - \frac{1}{3}\lambda R^3 = 0. \tag{6.6}$$

The first term of eq. (6.6) can be rewritten as

$$R^2[\ddot{R} + 2H_0\dot{R}] = \frac{R^2}{t}\frac{d^2}{dt^2}(Rt)$$

Multiplying eq. (6.6) by $(t/t_0)^3$ and setting $Y(t) = R(t)(t/t_0)$, one has :

$$Y^2\ddot{Y} - \frac{\lambda}{3}Y^3 + \frac{4\pi G\rho(t_0)}{3} = 0.$$

Multiplying this expression by $Y^2 dY/dt$ and integrating term by term, one finally obtains the Fridman equation for the scale function $Y(t)$ :

$$\dot{Y}^2 = \frac{8\pi G\rho(t_0)}{3Y} - k + \frac{\lambda}{3}Y^2. \tag{6.7}$$

Hence the substantial variation with respect to the usual Newtonian cosmology (that gives the same fundamental results as GR) is that the scale function is now given by $Y(t)$ instead of $R(t)$. In particular, $Y(t)$ satisfies the Fridman equation.
From eq. (6.7) one can see that the solution $R = 1$, $Y = R_0(t) = t/t_0$ corresponding to PSR exists only for an empty universe with $\rho$ and $\lambda$ null. Thus, the perfect cosmological principle is a special case of the usual cosmological principle which, however, is possible only for an empty universe.

The scale function that appears in the luminosity-distance law is now $Y$ instead of $R$. It is easy to see that this also holds for the expression of the redshift. From a Newtonian point of view, it is in fact acceptable to assume that the speed $c(t)$ of a light ray (which must depend only on $t$ due to the cosmological principle and which we will assume hereafter to be constant) may be composed with the expansion speed according to the law :

$$\frac{dx}{dt} = K(t)x \pm c = \frac{\dot{R}}{R}x + \frac{x}{t} \pm c = \frac{\dot{Y}}{Y}x \pm c$$

Which can be written as :

$$Y\frac{d}{dt}\left(\frac{x}{Y}\right) = \pm c .$$

Integrating this equation between the cosmic times $t_1$ and $t_2 > t_1$ one has :

$$\frac{x_2}{Y_2} - \frac{x_1}{Y_1} = \pm c \int_{t_1}^{t_2} \frac{dt}{Y(t)} \qquad (6.8)$$

Now let's consider two light rays emitted at $x_1$ respectively at the times $t_1$ and $t_1 + \Delta t_1$, received by an observer located at $x_2 = 0$ respectively at the times $t_2$ and $t_2 + \Delta t_2$. One has :

$$\frac{x_1(t_1)}{Y(t_1)} = c \int_{t_1}^{t_2} \frac{dt}{Y(t)} \quad ; \quad \frac{x_1(t_1 + \Delta t_1)}{Y(t_1 + \Delta t_1)} = c \int_{t_1 + \Delta t_1}^{t_2 + \Delta t_2} \frac{dt}{Y(t)} \quad ;$$

$$\frac{x_1(t_1 + \Delta t_1)}{Y(t_1 + \Delta t_1)} - \frac{x_1(t_1)}{Y(t_1)} = c \left[ \int_{t_1 + \Delta t_1}^{t_2 + \Delta t_2} \frac{dt}{Y(t)} - \int_{t_1}^{t_2} \frac{dt}{Y(t)} \right]$$

$$= c \left[ \int_{t_2}^{t_2 + \Delta t_2} \frac{dt}{Y(t)} - \int_{t_1}^{t_1 + \Delta t_1} \frac{dt}{Y(t)} \right] \approx c \left[ \frac{\Delta t_2}{Y(t_2)} - \frac{\Delta t_1}{Y(t_1)} \right] .$$

The ratio $x/Y$ is independent of $t$, since it coincides with the co-moving coordinate of the emitter; thus the expression obtained is null and we have :

$$\frac{\Delta t_2}{\Delta t_1} = \frac{Y(t_2)}{Y(t_1)}$$

Applying this relation to the period of the electromagnetic wave emitted and received, one obtains the final result :

$$Y(t) = R(t)\frac{t}{t_0} = \frac{1}{1+z} . \qquad (6.9)$$

Incidentally, we note that eq. (6.8) can be transformed into the corresponding equation of standard Newtonian cosmology :

$$\frac{x_2}{Y_2} - \frac{x_1}{Y_1} = \pm c \int_{t(\tau_1)}^{t(\tau_2)} \frac{d\tau}{R[t(\tau)]}$$

by regraduating the clock that measures the cosmic time in agreement with the Milne scale :

$$\tau = t_0 + t_0 \ln\left(\frac{t}{t_0}\right) \qquad (6.10)$$

The frequency of the light arriving to the observer will be $v_\tau$ according to the $\tau$-clock and $v_t$ according to the $t$-clock. Clearly $v_\tau = v_t (dt/d\tau) = v_t (t/t_0)$.
Let's suppose that we have a collection of extragalactic objects whose distances $x$ and redshifts $z$ are known. It is then possible to test a given solution $Y(t)$ of eq. (6.7) in the following manner. Using eq. (6.9) we obtain $t$ from the redshift $z$. From eq. (6.8) written in the form :

$$x = c \int_t^{t_0} \frac{d\sigma}{Y(\sigma)}$$

one obtains the distance $x$ when $Y=1$. The value of $x$ found in this manner must be compared with the experimental value; the parameters of the scale function can be varied so to optimize the agreement.
This analysis strategy has been applied to the data collection for 117 type Ia supernovae published by the SCP (Supernova Cosmology Project) and SNLS (Supernova Legacy Survey) collaborations. In particular, the data of tables 8 and 9 of ref. [31] have been analyzed, which also include the data reported in tables 1 and 2 of ref. [32]. The data of the SCP collection [32], which is definitive, concern 42 supernovae while the data of the SNLS collection [31] regard the first year of observation. Ref. [31] provides a distance module $\mu_B$ constructed starting from the apparent magnitude of the supernova evaluated in the rest frame of the supernova itself. The luminosity distance in parsec is obtained directly as $10^{[(\mu_B/5) + 1]}$, which we assume to be the uncorrected distance $x_{uncorr}$. Assuming a value of 19.6 x 10$^9$ ly for $r$, we apply eq. (3.9) so obtaining the correct distance $x_{corr}$ of the supernova.

We have chosen a $k = 0$ model because its consistency with the more recent results concerning the cosmic microwave background (CMB). The only $k = 0$ cosmological model able to fit the obtained curve is that with cosmological constant $\Lambda > 0$ :

$$Y(t) = \alpha\,[\sinh\,(\beta t)]^{2/3} \ . \tag{6.11}$$

The best least-square fit is respectively given by $\alpha = 0.359$, $\beta = 2.24(t_0)^{-1}$ [$r^2 = 0.963$], when the PSR distance correction [eq. (3.9)] is applied; $\alpha = 0.364$, $\beta = 2.22(t_0)^{-1}$ [$r^2 = 0.955$] otherwise.

From the relations [33] :

$$8\pi G\rho = (4/3)\,\alpha^3\beta^2\ , \qquad \Lambda c^2 = \lambda = (4/3)\,\beta^2\ , \tag{6.12}$$

where G is the Newtonian gravitation constant and $c$ is the speed of light, we then obtain the current density of matter in the universe $\rho$ and the cosmological constant $\Lambda$. The values obtained for these two quantities :

$$\rho = 4.7 \times 10^{-31}\ \mathrm{g\ cm}^{-3}\ , \qquad \Lambda = 1.9 \times 10^{-56}\ \mathrm{cm}^{-2} \tag{6.13}$$

are in good agreement with current estimations. The value of $\rho$ agrees, in order of magnitude, with all the data available in the literature from the nowadays historical estimation of Oort (1958) up to the recent results of WMAP. The value of $\Lambda$ is in perfect agreement with the most recent opinion that takes $\Lambda \approx 2 \times 10^{-56}\ h^2\ \mathrm{cm}^{-2}$.

The fact that $\Lambda \approx 1/t_0^2$ leads one to believe that the cosmological term is not an independent quantity, but that it is instead associated with the same group structure from which emerges $t_0$. Nevertheless the projective relativity does not provide any indication of a possible connection of this kind; if it really exists, its derivation seems to require a cosmological theory even more extensive, perhaps a quantum cosmology.

Fig. 3 shows the distance-redshift diagram of the 117 supernovae both before (blue points) and after (yellow points) the application of the PSR distance correction. The effect of the correction is clearly that of reducing the distances corresponding to high redshifts. The results of the respective fits are indicated by the lines of the same colour. We will not go further into an analysis of the residues, given the totally indicative and preliminary nature of this work. It is sufficient to consider that the distance correction applied [eq. (3.9)] is that corresponding to the PSR solution, and not to a solution with effective expansion of the space parameterized by $k = 0$, $\lambda > 0$, which is that subsequently used for the fitting. For a truly consistent discussion it would be necessary to have the distance correction relative to each cosmological model, which would in turn require a complete development of the gravitational theory based on eq. (1.1).

For comparison, Fig. 3 also shows the distance-redshift curve of the model without expansion with $R=1$, corresponding to PSR.

## 7. Conclusions

As a whole, the application of projective relativity in the cosmological ambit does not seem to lead to results that are in strong disagreement with observational data or usually accepted conceptions. Certainly, the applications presented here have been developed on the guidance of heuristic reasoning, since neither the solutions of the gravitational equations (1.1) nor a clear physical treatment, for example, of the propagation of the electromagnetic waves in PSR and PGR is currently available. As an example, equation (4.1) is certainly not valid if at large distances gravity and the cosmological constant modify the trajectory of the light rays, since in this case the spatial section is no longer Euclidean. It is for these reasons that we have omitted the comparison with other more widely known models in this article: it would be entirely premature. However, with all its limits and despite the numerous unresolved problems that are awaiting an answer, we believe that this preliminary analysis at least demonstrates the need to expand upon the argument and also indicates possible research directions.

## Acknowledgements

This work is dedicated to the memory of Prof. Giuseppe Arcidiacono, a truly exquisite person : his work will always remain a source of inspiration in this field. I would like to also thank my friend and colleague Ignazio Licata for support and ideas, and Erasmo Recami for useful discussions and indications.

## References

1. Fantappié L.; *Rend. Accad. Lincei* 17, fasc. 5 (1954)
2. Fantappié L.; *Collectanea Mathematica* XI, fasc. 2 (1959)
3. Arcidiacono G.; *Rend. Accad. Lincei* XVIII, fasc. 4 (1955)
4. Arcidiacono G.; *Rend. Accad. Lincei* XVIII, fasc. 5 (1955)
5. Arcidiacono G.; *Rend. Accad. Lincei* XVIII, fasc. 6 (1955)
6. Arcidiacono G.; *Rend. Accad. Lincei* XX, fasc. 4 (1956)
7. Arcidiacono G.; *Rend. Accad. Lincei* XX, fasc. 5 (1956)
8. Arcidiacono G.; *Collectanea Mathematica* X, 85-124 (1958)
9. Arcidiacono G.; *Collectanea Mathematica* XII, 3-23 (1960)
10. Arcidiacono G.; *Collectanea Mathematica* XIX, 51-71 (1968)
11. Arcidiacono G.; *Collectanea Mathematica* XX, 231-255 (1969)
12. Arcidiacono G.; *Collectanea Mathematica* XXIV, 1-25 (1973)
13. Arcidiacono G.; *Collectanea Mathematica* XXV, 295-317 (1974)
14. Arcidiacono G.; *Gen. Rel. Grav.* 7, 885-889 (1976)
15. Arcidiacono G.; *Gen. Rel. Grav.* 8, 865-870 (1977)
16. Arcidiacono G.; Projective relativity, Cosmology and Gravitation. Hadronic Press (USA), 1986
17. Arcidiacono G.; The theory of hyperspherical universes. International Center for Comparison and Synthesis, Rome, 1987
18. Arcidiacono G.; La teoria degli universi, vol. II. Di Renzo, Rome, 2000
19. Castelnuovo G.; *Rend. Accad. Lincei* XII, 263 (1930)
20. Castelnuovo G,: *Scientia* 40, 409 (1931)
21. Castelnuovo G.; *Mon. Not. Roy. Astron. Soc.* 91, 829 (1931)
22. Licata I.; *EJTP* 10, 211-224 (2006)
23. Licata I.; Osservando la Sfinge, cap. VI, VIII. Di Renzo, Rome, 2003
24. De Broglie L.; Théorie générale des particules a spin. Gauthier-Villars, Paris,1954
25. De Broglie L.; Une nouvelle conception de la lumière. Hermann, Paris,1934
26. Arcidiacono G.; *Collectanea Mathematica* XXV, 159-184 (1974)
27. Laurie J (ed.).; Cosmology now. BBC Publications, London, 1973
28. FIRST WWW Homepage (http://sundog.stsci.edu)
29. Hasinger G., Schartel N., Komossa S.; *Ap. J.* 573 (2), L77-L80 (2002)
30. Alcaniz J.S., Lima J.A.S., Cunha J.V.; *Mon. Not. Roy. Astron. Soc.* 340, L39 (2003)
31. SNLS Collaboration; astro-ph/0510447 v1 (2005)

32. Perlmutter, S., Aldering, G. et al. 1999, *Ap. J.*, 517, 565; astro-ph/9812133 v1 (1998)
33. Bondi H.; Cosmology. Cambridge University Press, Cambridge, 1961

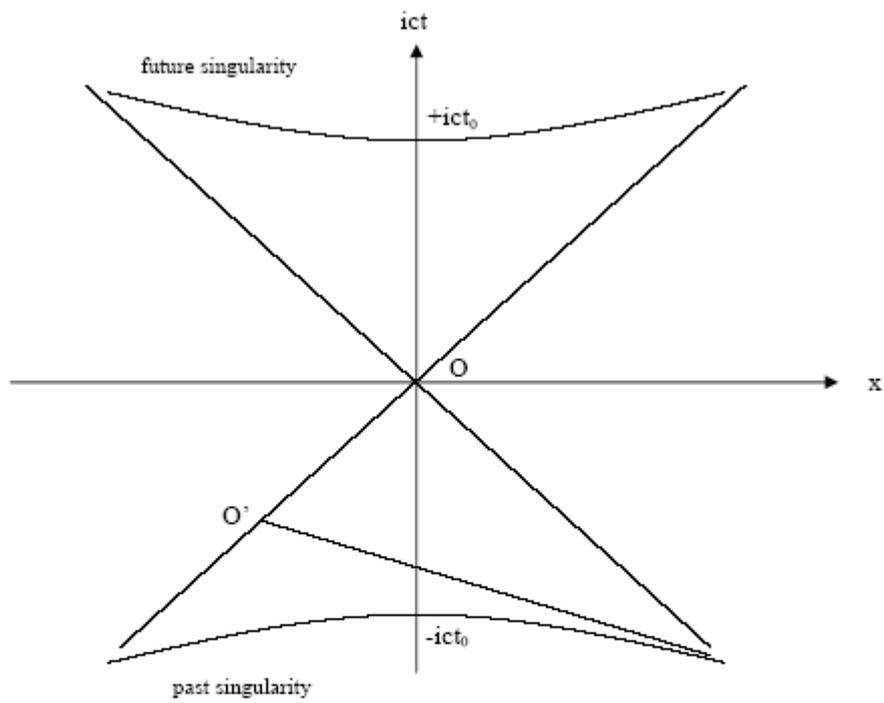

Fig. 1
The light cone with variable opening

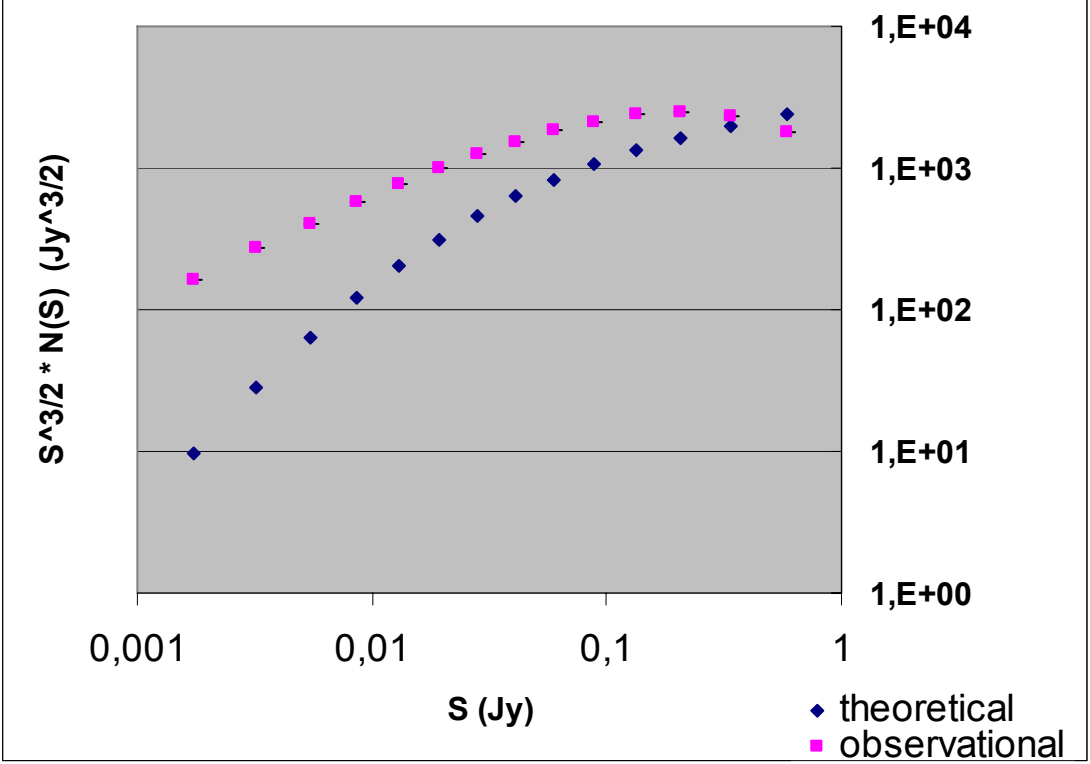

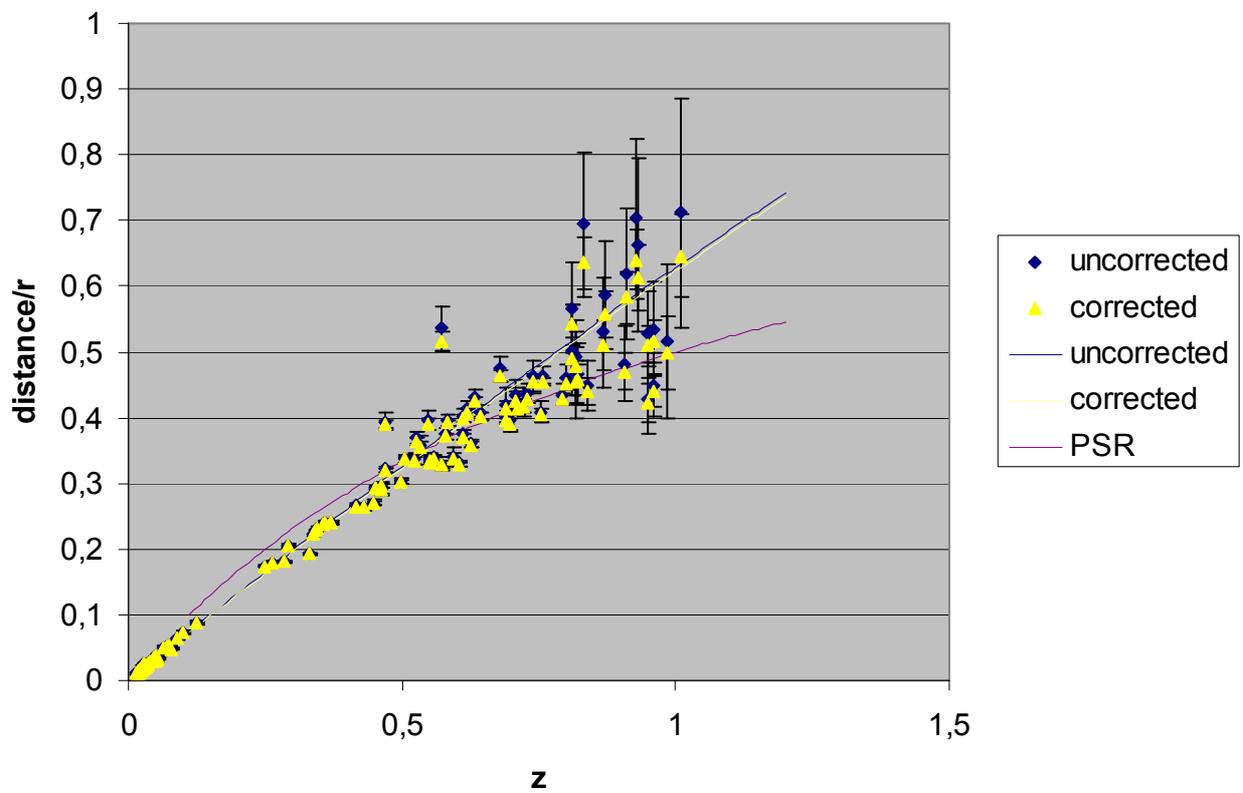
Fig. 3; Distance-z relation for type Ia supernovae